\documentclass{aa}  

\usepackage{txfonts,epsfig,graphicx,url,twoopt}
\usepackage[breaklinks=true]{hyperref} 
\usepackage[usenames, dvipsnames]{color}

\hypersetup{colorlinks=true,citecolor=blue,linkcolor=blue,urlcolor=red}
\graphicspath{{./figures/}}
\begin{document}

   \title{Planetesimal formation during protoplanetary disk buildup}
   \titlerunning{Planetesimal formation during disk buildup}

   \subtitle{}

   \author{J. Dr{\c a}{\. z}kowska\inst{1,2}
          \and
          C.~P. Dullemond\inst{3}
          }

   \institute{University of Zurich, Institute for Computational Science,
              Winterthurerstrasse 190, 8057 Zurich, Switzerland
              \and
              Current address: Max Planck Institute for Solar System Research, Justus-von-Liebig-Weg 3, 37077 G\"ottingen, Germany\\
              \email{drazkowska@mps.mpg.de}
           \and
              Heidelberg University, Center for Astronomy, Institute of Theoretical Astrophysics, Albert-Ueberle-Str.~2, 69120 Heidelberg, Germany
             }

   \date{Received -; accepted -}

\abstract
   {Models of dust coagulation and subsequent planetesimal formation are usually computed on the backdrop of an already fully formed protoplanetary disk model. At the same time, observational studies suggest that planetesimal formation should start early, possibly even before the protoplanetary disk is fully formed.}
   {In this paper we investigate under which conditions planetesimals already form during the disk buildup stage, in which gas and dust fall onto the disk from its parent molecular cloud.}
   {We couple our earlier planetesimal formation model at the water snow line to a simple model of disk formation and evolution.}
   {We find that there are parameters for which planetesimals already form during the disk buildup. This occurs when the viscosity driving the disk evolution is intermediate ($\alpha_{\mathrm{v}}\sim10^{-3}-10^{-2}$) while the turbulent mixing of the dust is reduced compared to that ($\alpha_{\rm t} \le 0.03 \cdot \alpha_{\rm v}$), and with the assumption that the water vapor is vertically well-mixed with the gas. Such a $\alpha_{\rm t}\ll\alpha_{\rm v}$ scenario could be expected for layered accretion, where the gas flow is mostly driven by the active surface layers, while the midplane layers, where most of the dust resides, are quiescent.}
   {In the standard picture where protoplanetary disk accretion is driven by global turbulence, we find that no planetesimals form during the disk buildup stage. Planetesimal formation during the buildup stage is only possible in scenarios in which pebbles reside in a quiescent midplane while the gas and water vapor are diffused at a higher rate.}

   \keywords{accretion, accretion disks -- 
                stars: circumstellar matter -- 
                protoplanetary disks -- 
                planet and satellites: formation -- 
                methods: numerical
               }

   \maketitle
%

{\bf This is an updated version of the original paper, including corrections described in the corrigendum: \url{https://doi.org/10.1051/0004-6361/201732221e}. We thank Raphael Marschall and Alessandro Morbidelli for their help with identifying the bug in the original code. The code used to generate results presented here is available at \url{https://doi.org/10.5281/zenodo.7657274
}. Always up-to-date version of the code is available at \url{https://github.com/astrojoanna/DD-diskevol}.}

\section{Introduction}

For reasons of convenience, most models of dust evolution and planetesimal formation in protoplanetary disks are computed against the backdrop of a simple stationary-state model for the gaseous disk \citep[see~e.g.,][]{1980Icar...44..172W, 1996A&A...309..301S,  1997A&A...319.1007S, 2008A&A...480..859B, 2012ApJ...752..106O, 2016A&A...594A.105D, 2017ApJ...839...16C,2017MNRAS.472.4117E}. At the start of the simulation the dust is assumed to be all in the form of micron-size monomers, and as time goes by, the monomers stick to each other and form ever larger dust aggregates. Whether it is justified to start the simulation with an already fully fledged class II protoplanetary disk, or if it is important to include the disk buildup stage, is a question that remains to be explored. The same question comes up when studying the meteoritic record: the ``time zero'' is defined as the time when calcium-aluminium-rich inclusions (CAIs) were formed, and all time is measured from that point onward. But is that definition of $t=0$ the end of the buildup phase or the beginning? It seems to be an important question, because during the buildup phase the disk is highly dynamic, possibly gravitationally unstable, and is continuously fed with fresh gas and dust.

This area has not been entirely ignored. Whether or not the later dust growth phase can be affected by events taking place during the buildup phase was investigated \citep{2008A&A...491..663D}. However, as was shown by \citet{2005A&A...434..971D} and  \citet{2010A&A...513A..79B}, dust growth quickly loses its ``memory'' of the initial conditions. If drift is not included, the coagulation and fragmentation equilibrium is quickly reached, and  which initial condition we start from is irrelevant. If we include radial drift, the situation changes slightly, to the extent that the dust may be depleted by radial drift slightly earlier or later, depending on the initial conditions. But overall, the initial conditions of the dust do not appear to be very important.

Things may change, however, if we include the formation of planetesimals. Studies of mass reservoirs of planet-forming disks infer that planet(esimal) formation should start early \citep{2010MNRAS.407.1981G, 2011MNRAS.412L..88G, 2014MNRAS.445.3315N, 2017ApJ...838..151T}. If we assume that the main mechanism responsible for planetesimal formation is the streaming instability \citep{2007Natur.448.1022J}, then the initial conditions may be extremely important as planetesimals can only form under particular conditions \citep{2009ApJ...704L..75J, 2015A&A...579A..43C}. These special conditions necessary for the streaming instability may be fleeting: a missed chance (by an unfavorable initial condition, for instance) could mean that a sufficiently dense population of sufficiently large dust aggregates does not form. As a consequence all the solids remain below the ``meter size barrier'', and no planetesimals are formed. It is therefore important to not simply start the model with an already formed disk filled with micron-sized monomers, but instead follow the formation of the disk and let dust grow and drift in the disk buildup stage.

The purpose of this paper is to investigate the influence of the class 0/I stage of the disk buildup on the first population of planetesimals. We follow the dust evolution and planetesimal formation model presented by \citet{2017A&A...608A..92D}, in which the growth of dust is computed, and, based on a simplified criterion for the triggering of the streaming instability, part of this dust is converted into planetesimals. In that paper it was found that processes taking place around the water snow line enable planetesimal formation. Most importantly, due to the fact that icy dust is more sticky, sufficiently large pebbles can grow outside of the snow line. Because the dry aggregates remain small and well-coupled to the gas, there is a ``traffic jam'' arising inside the snow line that slows down the removal of solids by radial drift. What is more, the outward diffusion of water vapor from inside the snow line followed by its recondensation (so called ``cold finger effect'') increases the abundance of icy pebbles that trigger planetesimal formation just outside of the snow line. In this paper, we include early phases of disk evolution in this picture.

This paper is organized as follows. We outline our numerical approach in Sect.~\ref{sub:methods}. We describe our results in Sect.~\ref{sub:results} and discuss their major limitations in Sect~\ref{sub:discussion}. Finally, we summarize our findings in Sect.~\ref{sub:conclusions}.

\section{Methods}\label{sub:methods}

\subsection{Gas disk}\label{sub:methodgas}
We choose the \citet{2005A&A...442..703H} approach, which solves the viscous disk equations supplemented with a mass source function that describes the disk buildup from a rotating collapsing molecular cloud core. The infalling cloud model is the inside-out \citet{1977ApJ...214..488S} model coupled to the \citet{1976ApJ...210..377U} rotating infall model. Our implementation was described in \citet{2006ApJ...640L..67D} and \citet{2006ApJ...645L..69D}. The disk model includes the angular momentum transport by gravitational instability, albeit in a local viscous disk approximation, which means that when the gravitational instability is detected with the Toomre criterion, the local gas viscosity is increased \citep[for details see][]{2001MNRAS.324..705A}. We found that neglecting the gravitational instability does not change our results significantly, as it only happens for a relatively short period of time and well outside of the planetesimal formation region.

The infalling cloud has an initial mass of $M_{\rm cloud}=1$~M$_{\odot}$, temperature of $T_0=10$~K, and rotates at a rate of $\Omega_{\rm 0} = 7\cdot10^{-15}$~s$^{-1}$. On a timescale of $\sim7\cdot10^{5}$~yr, this cloud forms a single star surrounded by a disk with a peak mass depending on the viscosity parameter $\alpha_{\rm v}$, but staying within the range of 0.1-0.2~M$_{\odot}$. 
We take into account heating due to viscosity and irradiation by the central star when calculating the midplane temperature in the disk. Our current model does not include photoevaporation, thus the disk lifetime is determined by its viscous evolution described by the standard $\alpha$-formalism \citep{1973A&A....24..337S}, where the gas viscosity is defined as
\begin{equation}\label{eq:gasvis}
\nu = \alpha_{\rm v} c_{\rm s} H_{\rm g},
\end{equation}
where $c_{\rm s}$ is the sound speed and $H_{\rm g}$ is the gas scale height, which is equal to the turbulent gas diffusivity $D_{\rm gas}$. Since we use a vertically averaged model, the diffusivity is essentially a density-weighted average and we cannot directly model the vertical distance at which the gas flow takes place.

We vary the disk viscosity parameter $\alpha_{\rm v}$ between $3\cdot10^{-4}$ and $10^{-2}$ and purposely distinguish it from $\alpha_{\rm t}$, which describes the strength of the midplane turbulence that affects vertical settling and fragmentation of dust aggregates. We consider models with $\alpha_{\rm t} \le \alpha_{\rm v}$, where the standard case of  $\alpha_{\rm t} = \alpha_{\rm v}$ is our benchmark model. The cases with $\alpha_{\rm t} < \alpha_{\rm v}$ are motivated by the recent protoplanetary disk models where a quiescent midplane layer is often found \citep{2014prpl.conf..411T}. We provide a more in-depth discussion of these values in Section~\ref{sub:discussion}.

\subsection{Dust evolution and planetesimal formation}
The infall of gas onto the disk is accompanied by the delivery of dust, with the usual dust-to-gas ratio of 1\%. We neglect dust coagulation inside the envelope, where the growth timescale is very long \citep{2009A&A...502..845O}. The infalling dust is assumed to be monomer size, and the coagulation in the disk at a given orbital distance $R$ only starts after the surface density of dust exceeds $\Sigma_{\rm{d, min}} = 10^{-6}$~g~cm$^{-2}$. The dust coagulation is modeled with the two-population algorithm proposed by \citet{2012A&A...539A.148B}. However, the initial growth stage is modified from the original algorithm to take into account the disk buildup stage when the mass of the central star and thus the rotation frequency $\Omega_{\rm K}(t,R) = \sqrt{GM_{\star}(t)/R^3}$ increase with time. We estimate the size in the initial growth regime as
\begin{equation}
a_{\rm{ini,i}} = a_{\rm{ini,i-1}}\cdot\exp\left({Z\cdot\Omega_{\rm K}\cdot\Delta t}\right),
\end{equation}
where $a_{\rm{ini,i-1}}$ is the maximum size obtained in the previous time step, $\Delta t$ is the time step duration, and $Z$ is the vertically integrated dust-to-gas ratio. The maximum aggregate size at each orbital distance is determined as a minimum of $a_{\rm{ini}}$, fragmentation limit $a_{\rm{frag}}$ and maximum size that can be retained taking into account the radial drift $a_{\rm{drift}}$.

We assume that the infalling dust consists of 50\% ice and 50\% rock. Inside the snow line the ice component sublimates and is added to the water vapor reservoir. We track the water vapor evolution and account for the possibility of its recondensation outside of the snow line.
Water sublimation and recondensation is included following the algorithm suggested by \citet{2017A&A...602A..21S}. Since the water ice is more sticky than silicate dust \citep{2009ApJ...702.1490W, 2011ApJ...737...36W, 2014MNRAS.437..690A}, we assume that aggregates outside of the snow line fragment at impact speeds above $v_{\rm f, out}=10$~m~s$^{-1}$ while inside the snow line the dry aggregates fragment as soon as the impact speed exceeds $v_{\rm f, in}=1$~m~s$^{-1}$. To avoid numerical oscillations, we use the following sigmoid function to model the transition between $v_{\rm f, out}$ and $v_{\rm f, in}$:
\begin{equation}
    v_{\rm{f}}(r) = 10\ \mathrm{m}\ \mathrm{s}^{-1}\ \slash\ \left(1 + 9\cdot \exp\left(-400\cdot\left(\frac{\Sigma_{\rm{ice}}}{\Sigma_{\rm{d}}}-0.01\right)\right)\right).
\end{equation}
This transition means that the fragmentation-limited size $a_{\rm{frag}}\propto v_{\rm f}^2$ is two orders of magnitude larger outside of the snow line than inside it. 

When calculating the radial drift velocity, we take into account the so-called collective drift effect, which means that the drift speed decreases as the solids-to-gas ratio increases.

Planetesimal formation is included in a simple way. We assume that planetesimals may be formed by streaming instability if the midplane dust-to-gas ratio calculated for the dust particles when size corresponding to the Stokes number ${\rm St}\ge10^{-2}$ exceeds unity. In every time step and at every orbital distance we verify whether this condition is fulfilled and if it is, we transfer part of the surface density of pebbles to planetesimals. Currently we do not include planetesimal evolution. 

We refer interested readers to \citet{2017A&A...608A..92D} for a more detailed discussion of our dust evolution and planetesimal formation treatment and its limitations.

\section{Results}\label{sub:results}

\begin{figure}
   \centering
   \includegraphics[width=0.9\hsize]{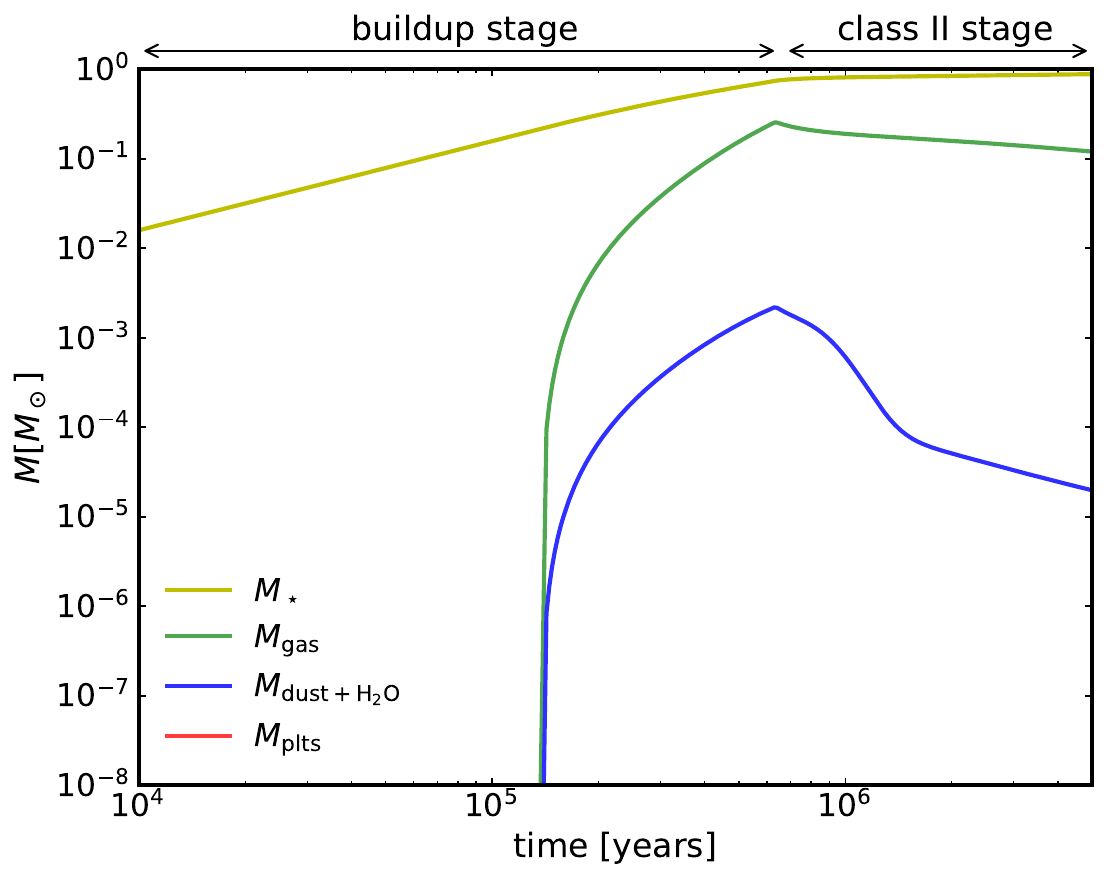}
      \caption{Time evolution of star mass, gas disk, dust and water mass in the benchmark model with $\alpha_{\rm v}=\alpha_{\rm t}=10^{-3}$.}
      \label{fig:massevo1}
\end{figure}

In a standard case, when the outward transport of angular momentum and thus disk accretion is driven by global, isotropic turbulence, $\alpha_{\rm v}$ and $\alpha_{\rm t}$ are equal. Following \citet{2017A&A...608A..92D}, we choose $\alpha_{\rm v}=\alpha_{\rm t}=10^{-3}$ for our benchmark run. Figure \ref{fig:massevo1} presents evolution of the total mass of the star, gas, dust disk, and planetesimals. Since the infall proceeds inside-out, matter only falls onto the star at first. After $\sim1.3\cdot10^{5}$~yr, the disk buildup starts and it lasts for another $\sim5\cdot10^{5}$~yr. After that, the class II disk stage starts and we model it for another 5~Myr. No planetesimals form in this model. This is because the disk created by the inside-out infall is too small to allow for a strong concentration of solids in the long term. The fact that a small disk size hinders planetesimal formation was already discussed in \citet[][their Sect. 3.3.2]{2017A&A...608A..92D}. 

Planetesimal formation requires the existence of a dense midplane layer of sufficiently large pebbles. We find that sufficiently large pebbles can only grow outside of the snow line, where the icy dust is more sticky. The highest dust-to-gas ratio is always obtained directly outside of the snow line and this is the region where planetesimals form. The mechanism of triggering this snow line pile-up is as follows: the large icy pebbles are efficiently delivered to the snow line because of their fast radial drift. The smaller aggregates inside the snow line drift at much lower speed and thus a pile-up arises, which spreads outside of the snow line by turbulent diffusion. The pile-up of icy pebbles outside of the snow line is supported by the cold finger effect and the collective drift, which is a key component for obtaining a sufficiently high pebble-to-gas ratio.

As explained above, there are two processes promoting the pileup of icy pebbles outside of the snow line in the protoplanetary disk: the outward diffusion and recondensation of water vapor (the cold finger effect) and the traffic-jam arising because the dry aggregates inside the snow line remain small and thus drift at a lower speed. The latter process operates on timescales of approximately 200,000 yr, during which pebbles grow in the outer disk and are transported inside the snow line by radial drift. This process can only start after the disk is fully formed. In the disk buildup stage, the effect of radial drift is limited, like in the case of an outburst \citep{2017A&A...605L...2S}. Thus, there is only the outward diffusion and recondensation of water vapor helping to form dust overdensity, so the dust-to-gas ratio enhancement outside of the snow line in this early stage is always weaker than the one arising during the following disk evolution, and in our benchmark run it is not sufficient to produce dense enough midplane layer of pebbles. 

\begin{figure}
   \centering
   \includegraphics[width=0.9\hsize]{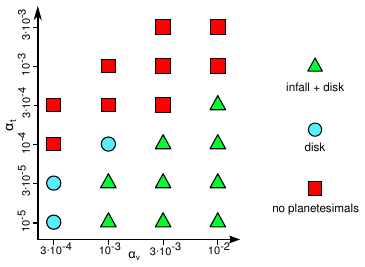}
      \caption{Summary of models showing the influence of the viscosity parameter $\alpha_{\rm v}$ and the midplane turbulence strength $\alpha_{\rm t}$. The triangles denote models where planetesimals are formed both during the disk buildup and in the protoplanetary disk stage. The circles mark models where planetesimals only form after the disk is fully formed, and squares mean no planetesimal formation at all.}
      \label{fig:alpha_alpha}
\end{figure}

Exploring the possibility of planetesimal formation during the disk buildup stage we found that it is mostly regulated by the viscosity parameter $\alpha_{\rm v}$ and the turbulence strength $\alpha_{\rm t}$. Figure~\ref{fig:alpha_alpha} gives the overview of results obtained in models with different $\alpha_{\rm v}$ and $\alpha_{\rm t}$. 

In general, we find that planetesimal formation is easier to trigger after the disk is fully formed than during the disk buildup. Each model that forms planetesimals in the disk buildup stage also forms them in the class II disk stage. Furthermore, planetesimals are formed in the protoplanetary disk stage in almost all the models with $\alpha_{\rm t} \le 10^{-4}$, even if there are no planetesimals formed during the disk buildup stage. This is because the density enhancement produced thanks to the traffic jam effect in the disk stage is significantly higher than the one given by the cold finger effect alone \citep[see][Sect. 3.2.2]{2017A&A...608A..92D}.
 
Nevertheless, we find that even the cold finger effect alone may be enough to form conditions for planetesimal formation. In a disk with a low internal turbulence level, a vertically integrated dust-to-gas ratio of 0.03 may be enough to trigger the streaming instability \citep{2010ApJ...722.1437B, 2014A&A...572A..78D, 2015A&A...579A..43C}. This kind of enhancement is easily produced by the outward diffusion and recondensation of water vapor \citep{1988Icar...75..146S, 2004ApJ...614..490C, 2017A&A...602A..21S}. The outward redistribution of water promoted by high $\alpha_{\rm v}$ values thus supports planetesimal formation particularly during the disk buildup stage, when it fully relies on the cold finger effect. At the same time however, the low midplane turbulence is necessary to allow for formation of large enough pebbles and a dense midplane layer, and consequently for planetesimal formation.

\begin{figure}
   \centering
   \includegraphics[width=0.9\hsize]{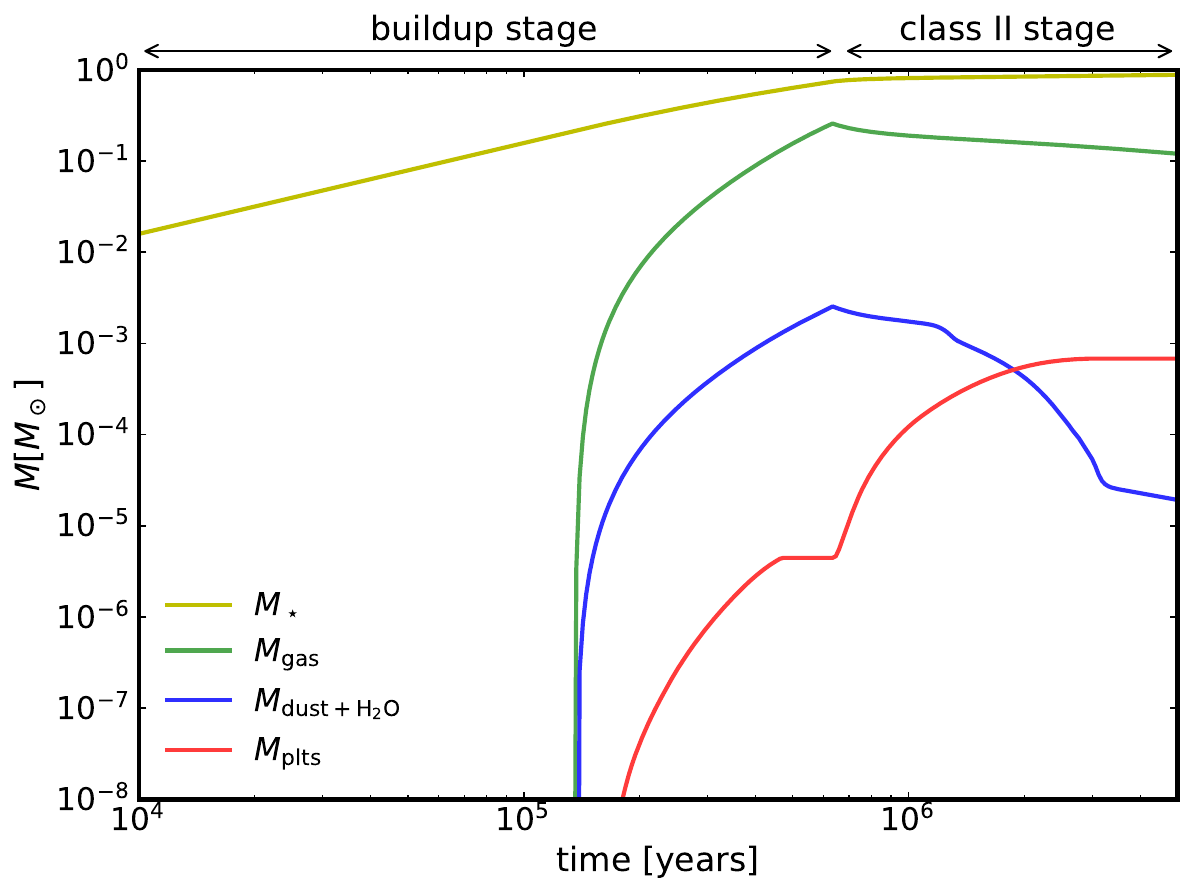}
      \caption{Time evolution of star mass, gas disk, dust and water, and planetesimal reservoir for model with $\alpha_{\rm v}=10^{-3}$ and $\alpha_{\rm t}=10^{-5}$.}
      \label{fig:massevo}
\end{figure}

\begin{figure}
   \centering
   \includegraphics[width=0.9\hsize]{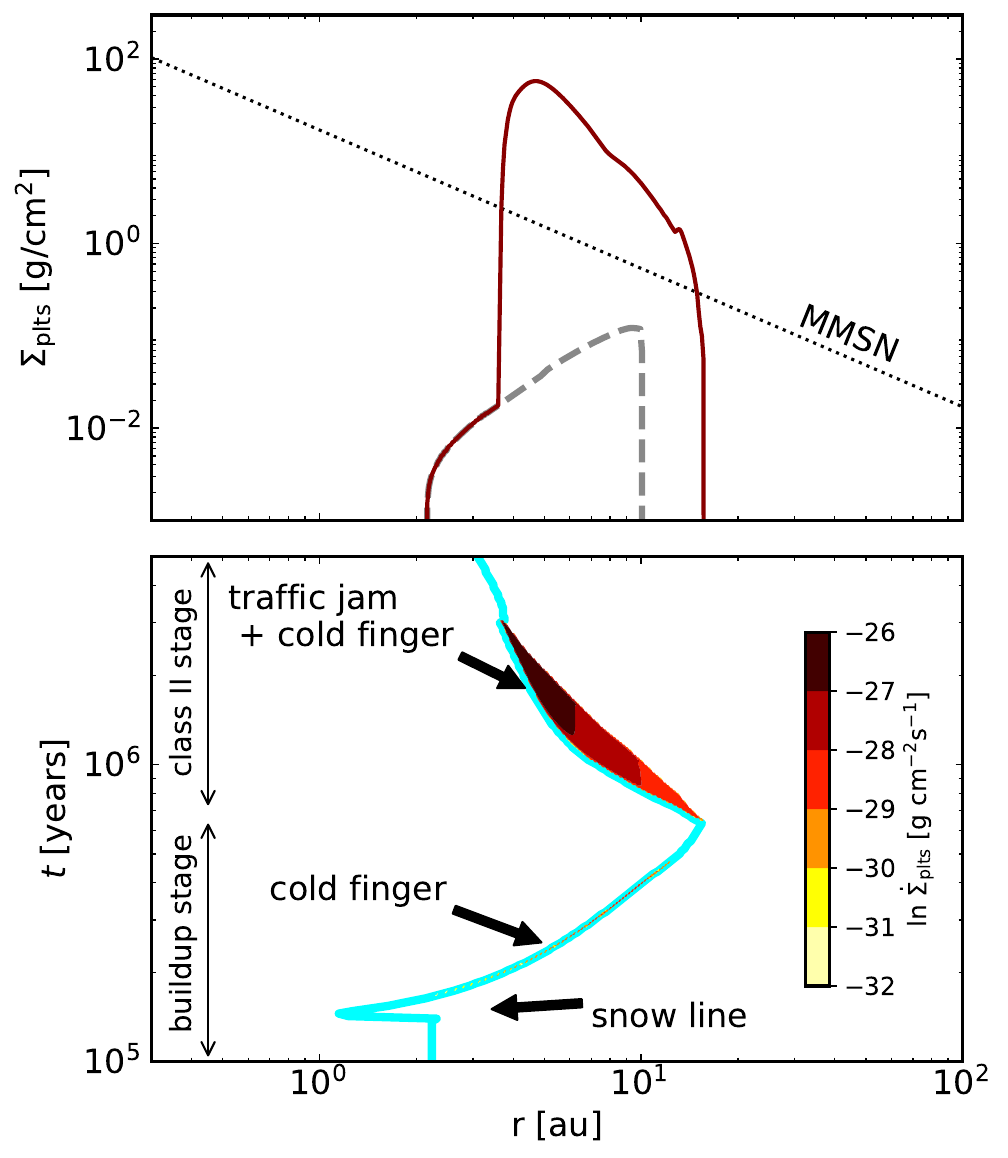}
      \caption{{\it Upper panel:} Surface density of planetesimals at the end of the disk buildup stage (at $7\cdot10^5$~years, gray dashed line) and at the end of the disk lifetime (at $10^7$~years, red solid line) obtained in the model with $\alpha_{\rm v}=10^{-3}$ and $\alpha_{\rm t}=10^{-5}$. The black dotted line corresponds to the minimum mass solar nebula. {\it Lower panel:} Radial and time distribution of planetesimal formation in the same model. The light blue solid line shows the location of the snow line.}
      \label{fig:plts}
\end{figure}

Figure \ref{fig:massevo} presents the evolution of the total mass of the star, gas and dust disk, and planetesimals in a model with $\alpha_{\rm v}=10^{-3}$ and $\alpha_{\rm t}=10^{-5}$. In contrast to the results presented in Fig.~\ref{fig:massevo1}, two periods of planetesimal formation are visible: first during the disk buildup about $1.5$~M$_{\oplus}$ of planetesimals is formed, and then in the more extended protoplanetary disk phase another $225$~M$_{\oplus}$ of planetesimals arises.

If at all possible, planetesimal formation in the disk buildup stage begins shortly after the region outside of the snow line is populated. Dust growth timescale can be estimated as 
\begin{equation}\label{eq:tgrowth}
\tau_{\rm growth} \approx \frac{1}{Z\cdot\Omega_{\rm K}} = 10^2\cdot\left(\frac{0.01}{Z}\right)\cdot\left(\frac{R}{1\ {\rm AU}}\right)^{1.5}\ {\rm yr}
,\end{equation}
\citep[see][]{2012A&A...539A.148B}, which, for a typical location of the water snow line, is much shorter than the buildup and disk evolution timescales. Derivation of Eq.~(\ref{eq:tgrowth}) assumes that collisions are driven by turbulence. In cases of very low $\alpha_{\rm t}$, collisions could also be driven by differential drift. In that case, analogical derivation shows that the growth is slowed down by a factor of $c_{\rm s} / v_\eta$, where $v_\eta$ is the maximum drift velocity. This factor is on the order of 20 in a typical protoplanetary disk. Thus, pebbles can easily grow to their maximum size even before the disk is fully formed. The timescale for outward diffusion and recondensation of water vapor is also short. The condensation is assumed to be instantaneous since the abundance of small grains is replenished by fragmentation \citep[for more discussion see][]{2017A&A...608A..92D}. 
The timescale for diffusion can be estimated as
\begin{equation}\label{eq:tdiff}
\tau_{\rm{diff}} \approx \frac{\Delta R^2}{D_{\rm{gas}}} = 10^3\cdot\left(\frac{10^{-3}}{\alpha_{\rm v}}\right)\cdot\left(\frac{R}{1\ {\rm AU}}\right)^{1.5}\cdot\left(\frac{\Delta R}{H_{\rm g}}\right)^2\ {\rm yr},
\end{equation}
where $\Delta R$ is the radial width of the recondensation region and gas diffusivity $D_{\rm{gas}}$ is equal to gas viscosity given in Eq.~(\ref{eq:gasvis}). Such a short timescale allows for a moderate enhancement of the dust-to-gas ratio outside of the snow line to appear even at the very early stages of disk buildup, which leads to planetesimal formation in some of the models.

Figure~\ref{fig:plts} presents the distribution of planetesimals formed during the buildup and the fully formed disk stage in a model with $\alpha_{\rm v}=10^{-3}$ and $\alpha_{\rm t}=10^{-5}$. Planetesimal formation follows the water snow line. This is initially close-in, and the planetesimal formation in the disk buildup stage starts already at 2~AU. Then the snow line gradually moves outwards as the disk becomes more massive and heats up. Since both $\tau_{\rm growth}$ and $\tau_{\rm{diff}}$ that determine the efficiency of the cold finger effect depend strongly on the distance from the star (Eqs~(\ref{eq:tgrowth})-(\ref{eq:tdiff})), no planetesimals are formed anymore when the snow line recedes beyond $\sim$10~AU. In the class II disk phase, the snow line moves back inwards as the disk cools down and the more extended planetesimal formation phase, supported by the radial drift, takes place.
In the disk buildup stage, planetesimals form only directly outside of the snow line, and the planetesimal formation region at a given time is relatively narrow, corresponding to the width of the recondensation zone. After the disk is fully formed, the planetesimal formation region is wider thanks to the collective drift effect, which spreads the peak of the dust-to-gas ratio enhancement.

The surface density of planetesimals displayed in the upper panel of Fig.~\ref{fig:plts} is significantly higher than the conventional minimum mass solar nebula profile of \citet{1977Ap&SS..51..153W}, which means that the amount of planetesimals formed is sufficient to form the solar system planets. The mass of planetesimals formed during the buildup stage (the gray dashed line) is typically only a small fraction of the final planetesimal mass. Taking into account all the runs presented in Fig.~\ref{fig:alpha_alpha}, it ranges from 0 to $87\%$, but the higher values are only obtained when the total planetesial production is low ($\sim$5 M$_\oplus$). However, from the perspective of planet formation, these first planetesimals may be significant and trigger the early formation of planetary embryos.

\section{Discussion}\label{sub:discussion}

Section 2.3 of \citet{2017A&A...608A..92D} presented a comprehensive list of the limitations of our dust evolution algorithm. In this Section, we only address the aspects that are particularly important in the context of models presented in this paper.

\subsection{$\alpha_{\rm v}$ and $\alpha_{\rm t}$ values}

As summarized in Fig.~\ref{fig:alpha_alpha}, the values for the $\alpha_{\rm v}$, describing the global efficiency of angular momentum transport via turbulent viscosity, and the midplane turbulence strength parameter $\alpha_{\rm t}$ are crucial for the possibility of planetesimal formation in the buildup stage of protoplanetary disk. Thus, we  discuss a realistic range of these parameters below.

As mentioned in Sect.~\ref{sub:methodgas}, we use a vertically averaged method and thus we are not able to model the gas and dust flow at different layers directly. Thus, the $\alpha_{\rm v}$ describes density-averaged flow of gas (and water vapor) through the disk, which can be directly translated into disk lifetime. At the same time, $\alpha_{\rm t}$ concerns only the turbulence (i.e.,~the velocity dispersion) in the midplane, where the pebbles reside, and is used to constrain the scale height of the pebble layers as well as their collision speeds. 

Observational constraints on protoplanetary disk lifetimes are in the range of 1 to 10~Myr (\citealt{2007ApJ...662.1067H}, however \citealt{2014ApJ...793L..34P} pointed out that these short lifetimes may be a selection bias). This suggests values of $\alpha_{\rm v}$ close to 10$^{-2}$ if we assume that the dispersal is driven solely by gas accretion onto the central star. However, if the disk dispersal is also driven by other processes, such as magnetic winds and photoevaporation, the $\alpha_{\rm v}$ value can be lower. Thus, we considered $3\cdot10^{-4}\le\alpha_{\rm v}\le10^{-2}$.

Attempts to observationally constrain the disk turbulence brought various results, from $\alpha_{\rm t}<10^{-3}$ for the outer parts of the disk around HD~163296 \citep{2015ApJ...813...99F, 2017ApJ...843..150F} to $\alpha_{\rm t}\approx10^{-2}$ in the outer parts of the TW~Hya disk \citep{2016A&A...592A..49T}, both derived from molecular line emission. Interestingly, \citet{2016ApJ...816...25P} found the best fit between observations and models of the disk around HL~Tau assuming $\alpha_{\rm t} = 3\cdot10^{-4}$ as their dust settling parameter. In our paper, we considered $10^{-5}\le\alpha_{\rm t}\le3\cdot10^{-3}$. Although in principle
we treat $\alpha_{\rm v}$ and $\alpha_{\rm t}$ as independent parameters, we excluded the cases with $\alpha_{\rm t} > \alpha_{\rm v}$.

In a standard understanding of the layered accretion, which was based on including the Ohmic resistivity into the picture of magnetorotational instability, the flow of gas takes place in the active, upper layers of the disk while the midplane is ``dead'', with little or no turbulence \citep{1996ApJ...457..355G}. However, more recent models show that while turbulence measured by the local velocity dispersion, may be indeed many orders of magnitude lower in the midplane than in the surface layers, the gas density of the active surface layer is low and thus the density-weighted average of the turbulent diffusivity is not much higher than the midplane diffusivity \citep{2011ApJ...742...65O, 2018ApJ...865...10S}. This would suggest that the values of $\alpha_{\rm v}$ and $\alpha_{\rm t}$ should not be independent, and based on the results of \citet{2011ApJ...742...65O}, $\alpha_{\rm t}$ should be about ten times lower than $\alpha_{\rm v}$.

We would like to note that including non-ideal magnetohydrodynamic effects other than the Ohmic resistivity has recently changed the picture of protoplanetary disk evolution \citep{2016ApJ...821...80B, 2018ApJ...865...10S}. In general, it was found that large regions of the protoplanetary disk can be free of turbulence and the angular momentum can be removed vertically by magnetic wind \citep{2014ApJ...791..137B, 2017ApJ...845...75B} or radially through laminar torques \citep{2014A&A...566A..56L}. The consequences of this new protoplanetary disk picture for planetesimal formation are yet to be studied.

For reference, in planetesimal formation models similar to ours, \citet{2017ApJ...839...16C} used $\alpha_{\rm v}=10^{-2}$ and $\alpha_{\rm t}=10^{-4}$, while \citet{2017MNRAS.472.4117E} used $\alpha_{\rm v}=7\cdot10^{-4}$ and $\alpha_{\rm t}=7\cdot10^{-6}$.

\subsection{Vertical mixing}

Our model assumes that the water vapor is always well mixed with the gas and has the same scale height. Thus the advection and diffusion of water vapor is governed by the same diffusion coefficient as the gas, which is the $\alpha_{\rm v}$. However, if the water vapor is released by pebbles that reside in a thin midplane layer, which is regulated by $\alpha_{\rm t}$, the vertical diffusion of vapor should take place over timescales on the order of $\alpha_{\rm v}/\alpha_{\rm t}$ longer than radial diffusion (see Eq.~\ref{eq:tdiff}). In the runs with $\alpha_{\rm t}\ll\alpha_{\rm v}$, which actually produce planetesimals during the disk buildup stage, this could potentially limit the efficiency of the cold finger effect. 

However, in the class II protoplanetary disk stage, the cold finger effect impact on creating the pebble pileup outside of the snow line is secondary in comparison to the traffic jam effect caused by different sticking properties of icy and dry aggregates \citep[see][Sect 3.2.2]{2017A&A...608A..92D}. Thus we argue that the fact of neglecting the timescale of vertical mixing of water vapor does not impact our results in the disk phase. A similar conclusion was made by \citet[][(their Sect. 4.2.3)]{2017A&A...602A..21S}, who compared models with the scale height of water vapor equal to both gas and pebbles.

On the other hand, during the disk buildup stage the water vapor is primarily not delivered by the drifting pebbles but by the small icy dust grains falling onto the disk surface from the cold molecular cloud. Since evaporation of these monomer-sized grains is virtually instantaneous, the vapor is delivered directly to the active surface layer and the diffusion of vapor is indeed governed by the viscosity parameter $\alpha_{\rm v}$. What may be problematic is the vertical mixing of vapor outside the snow line, such that it increases the midplane pebble-to-gas ratio. This process might be potentially sped up by the sedimentation-driven coagulation \citep[see e.g.,][]{2005A&A...434..971D} if the water vapor freezes on small grains present in the upper layers that stick together and settle to the midplane. Detailed quantification of this process requires performing two-dimensional models, which is beyond the scope of this paper.

\section{Conclusions}\label{sub:conclusions}

We present the first study on planetesimal formation during the protoplanetary disk formation stage. Our key findings may be summarized as follows:
\begin{itemize}
\item{The water snow line is the preferable place for planetesimal formation, both when the disk is already fully formed and during its buildup.}
\item{Planetesimal formation is less likely to take place during the disk buildup phase than during the class II protoplanetary disk stage; it is only possible if the gas and water vapor redistribution is efficient ($\alpha_{\rm v} \ge 10^{-3}$) and the dust resides in a quiescent midplane ($\alpha_{\rm t} \le 0.03 \cdot \alpha_{\rm v}$). At the same time, the water vapor needs to be efficiently mixed to the gas scale height, which might be a condition that is incompatible with the required low $\alpha_{\rm t}$. The plausibility of such a setup in a realistic protoplanetary disk is unclear.}
\item{If planetesimals are already formed in the disk buildup stage, their mass is typically low, on the order of a single Earth mass, but due to their early occurrence, these bodies may be important for planet formation. }
\end{itemize}

Our findings may have implications for the formation history of meteorite parent bodies in the solar system, since early formation would lead to strong planetesimal melting by $^{26}$Al decay, while the later-formed planetesimals may be spared such complete melting \citep{1993Sci...259..653G, 2006M&PS...41...95H, 2016Icar..274..350L}. In fact, there is evidence that the differentiated parent bodies of iron meteorites formed earlier than the chondrite parent bodies \citep{2003Natur.422..502T, 2005ApJ...632L..41B}. Planetesimals formed in our models are icy, as they always form outside of the snow line \citep{2017A&A...608A..92D}. However, as the snow line location evolves, some of the planetesimals that are formed during the disk buildup stage are placed inside of it in the protoplanetary disk stage. Although there are still numerous intermediate stages involved in turning planetesimals into planets, which are beyond the scope of this work, our findings may certainly have implications for the final planetary architectures and compositions, because the question of whether planetesimals form early or late will have repercussions on which kind of planet may form, its position, and when.

\begin{acknowledgements}
We want to thank the referee, Satoshi Okuzumi, for his thoughtful report that helped us to clarify this paper. 
This work has been carried out within the framework of the National Centre for Competence in Research PlanetS supported by the Swiss National Science Foundation. JD acknowledges the financial support of the SNSF. CPD thanks the NCCR PlanetS for the hospitality during the month of July 2017, when this project was initiated.
\end{acknowledgements}

\bibliographystyle{aa} 
\bibliography{infall.bib}

\begin{thebibliography}{57}
\expandafter\ifx\csname natexlab\endcsname\relax\def\natexlab#1{#1}\fi

\bibitem[{{Armitage} {et~al.}(2001){Armitage}, {Livio}, \&
  {Pringle}}]{2001MNRAS.324..705A}
{Armitage}, P.~J., {Livio}, M., \& {Pringle}, J.~E. 2001, \mnras, 324, 705

\bibitem[{{Aumatell} \& {Wurm}(2014)}]{2014MNRAS.437..690A}
{Aumatell}, G. \& {Wurm}, G. 2014, \mnras, 437, 690

\bibitem[{{Bai}(2014)}]{2014ApJ...791..137B}
{Bai}, X.-N. 2014, \apj, 791, 137

\bibitem[{{Bai}(2016)}]{2016ApJ...821...80B}
{Bai}, X.-N. 2016, \apj, 821, 80

\bibitem[{{Bai}(2017)}]{2017ApJ...845...75B}
{Bai}, X.-N. 2017, \apj, 845, 75

\bibitem[{{Bai} \& {Stone}(2010)}]{2010ApJ...722.1437B}
{Bai}, X.-N. \& {Stone}, J.~M. 2010, \apj, 722, 1437

\bibitem[{{Birnstiel} {et~al.}(2010){Birnstiel}, {Dullemond}, \&
  {Brauer}}]{2010A&A...513A..79B}
{Birnstiel}, T., {Dullemond}, C.~P., \& {Brauer}, F. 2010, \aap, 513, A79

\bibitem[{{Birnstiel} {et~al.}(2012){Birnstiel}, {Klahr}, \&
  {Ercolano}}]{2012A&A...539A.148B}
{Birnstiel}, T., {Klahr}, H., \& {Ercolano}, B. 2012, \aap, 539, A148

\bibitem[{{Bizzarro} {et~al.}(2005){Bizzarro}, {Baker}, {Haack}, \&
  {Lundgaard}}]{2005ApJ...632L..41B}
{Bizzarro}, M., {Baker}, J.~A., {Haack}, H., \& {Lundgaard}, K.~L. 2005, \apjl,
  632, L41

\bibitem[{{Brauer} {et~al.}(2008){Brauer}, {Dullemond}, \&
  {Henning}}]{2008A&A...480..859B}
{Brauer}, F., {Dullemond}, C.~P., \& {Henning}, T. 2008, \aap, 480, 859

\bibitem[{{Carrera} {et~al.}(2017){Carrera}, {Gorti}, {Johansen}, \&
  {Davies}}]{2017ApJ...839...16C}
{Carrera}, D., {Gorti}, U., {Johansen}, A., \& {Davies}, M.~B. 2017, \apj, 839,
  16

\bibitem[{{Carrera} {et~al.}(2015){Carrera}, {Johansen}, \&
  {Davies}}]{2015A&A...579A..43C}
{Carrera}, D., {Johansen}, A., \& {Davies}, M.~B. 2015, \aap, 579, A43

\bibitem[{{Cuzzi} \& {Zahnle}(2004)}]{2004ApJ...614..490C}
{Cuzzi}, J.~N. \& {Zahnle}, K.~J. 2004, \apj, 614, 490

\bibitem[{{Dominik} \& {Dullemond}(2008)}]{2008A&A...491..663D}
{Dominik}, C. \& {Dullemond}, C.~P. 2008, \aap, 491, 663

\bibitem[{{Dr{\c a}{\.z}kowska} \& {Alibert}(2017)}]{2017A&A...608A..92D}
{Dr{\c a}{\.z}kowska}, J. \& {Alibert}, Y. 2017, \aap, 608, A92

\bibitem[{{Dr{\c a}{\.z}kowska} {et~al.}(2016){Dr{\c a}{\.z}kowska}, {Alibert},
  \& {Moore}}]{2016A&A...594A.105D}
{Dr{\c a}{\.z}kowska}, J., {Alibert}, Y., \& {Moore}, B. 2016, \aap, 594, A105

\bibitem[{{Dr{\c a}{\.z}kowska} \& {Dullemond}(2014)}]{2014A&A...572A..78D}
{Dr{\c a}{\.z}kowska}, J. \& {Dullemond}, C.~P. 2014, \aap, 572, A78

\bibitem[{{Dullemond} {et~al.}(2006{\natexlab{a}}){Dullemond}, {Apai}, \&
  {Walch}}]{2006ApJ...640L..67D}
{Dullemond}, C.~P., {Apai}, D., \& {Walch}, S. 2006{\natexlab{a}}, \apjl, 640,
  L67

\bibitem[{{Dullemond} \& {Dominik}(2005)}]{2005A&A...434..971D}
{Dullemond}, C.~P. \& {Dominik}, C. 2005, \aap, 434, 971

\bibitem[{{Dullemond} {et~al.}(2006{\natexlab{b}}){Dullemond}, {Natta}, \&
  {Testi}}]{2006ApJ...645L..69D}
{Dullemond}, C.~P., {Natta}, A., \& {Testi}, L. 2006{\natexlab{b}}, \apjl, 645,
  L69

\bibitem[{{Ercolano} {et~al.}(2017){Ercolano}, {Jennings}, {Rosotti}, \&
  {Birnstiel}}]{2017MNRAS.472.4117E}
{Ercolano}, B., {Jennings}, J., {Rosotti}, G., \& {Birnstiel}, T. 2017, \mnras,
  472, 4117

\bibitem[{{Flaherty} {et~al.}(2017){Flaherty}, {Hughes}, {Rose}, {Simon}, {Qi},
  {Andrews}, {K{\'o}sp{\'a}l}, {Wilner}, {Chiang}, {Armitage}, \&
  {Bai}}]{2017ApJ...843..150F}
{Flaherty}, K.~M., {Hughes}, A.~M., {Rose}, S.~C., {et~al.} 2017, \apj, 843,
  150

\bibitem[{{Flaherty} {et~al.}(2015){Flaherty}, {Hughes}, {Rosenfeld},
  {Andrews}, {Chiang}, {Simon}, {Kerzner}, \& {Wilner}}]{2015ApJ...813...99F}
{Flaherty}, K.~M., {Hughes}, A.~M., {Rosenfeld}, K.~A., {et~al.} 2015, \apj,
  813, 99

\bibitem[{{Gammie}(1996)}]{1996ApJ...457..355G}
{Gammie}, C.~F. 1996, \apj, 457, 355

\bibitem[{{Greaves} \& {Rice}(2010)}]{2010MNRAS.407.1981G}
{Greaves}, J.~S. \& {Rice}, W.~K.~M. 2010, \mnras, 407, 1981

\bibitem[{{Greaves} \& {Rice}(2011)}]{2011MNRAS.412L..88G}
{Greaves}, J.~S. \& {Rice}, W.~K.~M. 2011, \mnras, 412, L88

\bibitem[{{Grimm} \& {McSween}(1993)}]{1993Sci...259..653G}
{Grimm}, R.~E. \& {McSween}, H.~Y. 1993, Science, 259, 653

\bibitem[{{Hern{\'a}ndez} {et~al.}(2007){Hern{\'a}ndez}, {Hartmann}, {Megeath},
  {Gutermuth}, {Muzerolle}, {Calvet}, {Vivas}, {Brice{\~n}o}, {Allen},
  {Stauffer}, {Young}, \& {Fazio}}]{2007ApJ...662.1067H}
{Hern{\'a}ndez}, J., {Hartmann}, L., {Megeath}, T., {et~al.} 2007, \apj, 662,
  1067

\bibitem[{{Hevey} \& {Sanders}(2006)}]{2006M&PS...41...95H}
{Hevey}, P.~J. \& {Sanders}, I.~S. 2006, Meteoritics and Planetary Science, 41,
  95

\bibitem[{{Hueso} \& {Guillot}(2005)}]{2005A&A...442..703H}
{Hueso}, R. \& {Guillot}, T. 2005, \aap, 442, 703

\bibitem[{{Johansen} {et~al.}(2007){Johansen}, {Oishi}, {Mac Low}, {Klahr},
  {Henning}, \& {Youdin}}]{2007Natur.448.1022J}
{Johansen}, A., {Oishi}, J.~S., {Mac Low}, M.-M., {et~al.} 2007, \nat, 448,
  1022

\bibitem[{{Johansen} {et~al.}(2009){Johansen}, {Youdin}, \& {Mac
  Low}}]{2009ApJ...704L..75J}
{Johansen}, A., {Youdin}, A., \& {Mac Low}, M.-M. 2009, \apjl, 704, L75

\bibitem[{{Lesur} {et~al.}(2014){Lesur}, {Kunz}, \&
  {Fromang}}]{2014A&A...566A..56L}
{Lesur}, G., {Kunz}, M.~W., \& {Fromang}, S. 2014, \aap, 566, A56

\bibitem[{{Lichtenberg} {et~al.}(2016){Lichtenberg}, {Golabek}, {Gerya}, \&
  {Meyer}}]{2016Icar..274..350L}
{Lichtenberg}, T., {Golabek}, G.~J., {Gerya}, T.~V., \& {Meyer}, M.~R. 2016,
  \icarus, 274, 350

\bibitem[{{Najita} \& {Kenyon}(2014)}]{2014MNRAS.445.3315N}
{Najita}, J.~R. \& {Kenyon}, S.~J. 2014, \mnras, 445, 3315

\bibitem[{{Okuzumi} \& {Hirose}(2011)}]{2011ApJ...742...65O}
{Okuzumi}, S. \& {Hirose}, S. 2011, \apj, 742, 65

\bibitem[{{Okuzumi} {et~al.}(2012){Okuzumi}, {Tanaka}, {Kobayashi}, \&
  {Wada}}]{2012ApJ...752..106O}
{Okuzumi}, S., {Tanaka}, H., {Kobayashi}, H., \& {Wada}, K. 2012, \apj, 752,
  106

\bibitem[{{Ormel} {et~al.}(2009){Ormel}, {Paszun}, {Dominik}, \&
  {Tielens}}]{2009A&A...502..845O}
{Ormel}, C.~W., {Paszun}, D., {Dominik}, C., \& {Tielens}, A.~G.~G.~M. 2009,
  \aap, 502, 845

\bibitem[{{Pfalzner} {et~al.}(2014){Pfalzner}, {Steinhausen}, \&
  {Menten}}]{2014ApJ...793L..34P}
{Pfalzner}, S., {Steinhausen}, M., \& {Menten}, K. 2014, \apjl, 793, L34

\bibitem[{{Pinte} {et~al.}(2016){Pinte}, {Dent}, {M{\'e}nard}, {Hales}, {Hill},
  {Cortes}, \& {de Gregorio-Monsalvo}}]{2016ApJ...816...25P}
{Pinte}, C., {Dent}, W.~R.~F., {M{\'e}nard}, F., {et~al.} 2016, \apj, 816, 25

\bibitem[{{Schoonenberg} {et~al.}(2017){Schoonenberg}, {Okuzumi}, \&
  {Ormel}}]{2017A&A...605L...2S}
{Schoonenberg}, D., {Okuzumi}, S., \& {Ormel}, C.~W. 2017, \aap, 605, L2

\bibitem[{{Schoonenberg} \& {Ormel}(2017)}]{2017A&A...602A..21S}
{Schoonenberg}, D. \& {Ormel}, C.~W. 2017, \aap, 602, A21

\bibitem[{{Shakura} \& {Sunyaev}(1973)}]{1973A&A....24..337S}
{Shakura}, N.~I. \& {Sunyaev}, R.~A. 1973, \aap, 24, 337

\bibitem[{{Shu}(1977)}]{1977ApJ...214..488S}
{Shu}, F.~H. 1977, \apj, 214, 488

\bibitem[{{Simon} {et~al.}(2018){Simon}, {Bai}, {Flaherty}, \&
  {Hughes}}]{2018ApJ...865...10S}
{Simon}, J.~B., {Bai}, X.-N., {Flaherty}, K.~M., \& {Hughes}, A.~M. 2018, \apj,
  865, 10

\bibitem[{{Stepinski} \& {Valageas}(1996)}]{1996A&A...309..301S}
{Stepinski}, T.~F. \& {Valageas}, P. 1996, \aap, 309, 301

\bibitem[{{Stepinski} \& {Valageas}(1997)}]{1997A&A...319.1007S}
{Stepinski}, T.~F. \& {Valageas}, P. 1997, \aap, 319, 1007

\bibitem[{{Stevenson} \& {Lunine}(1988)}]{1988Icar...75..146S}
{Stevenson}, D.~J. \& {Lunine}, J.~I. 1988, \icarus, 75, 146

\bibitem[{{Teague} {et~al.}(2016){Teague}, {Guilloteau}, {Semenov}, {Henning},
  {Dutrey}, {Pi{\'e}tu}, {Birnstiel}, {Chapillon}, {Hollenbach}, \&
  {Gorti}}]{2016A&A...592A..49T}
{Teague}, R., {Guilloteau}, S., {Semenov}, D., {et~al.} 2016, \aap, 592, A49

\bibitem[{{Trieloff} {et~al.}(2003){Trieloff}, {Jessberger}, {Herrwerth},
  {Hopp}, {Fi{\'e}ni}, {Gh{\'e}lis}, {Bourot-Denise}, \&
  {Pellas}}]{2003Natur.422..502T}
{Trieloff}, M., {Jessberger}, E.~K., {Herrwerth}, I., {et~al.} 2003, \nat, 422,
  502

\bibitem[{{Tsukamoto} {et~al.}(2017){Tsukamoto}, {Okuzumi}, \&
  {Kataoka}}]{2017ApJ...838..151T}
{Tsukamoto}, Y., {Okuzumi}, S., \& {Kataoka}, A. 2017, \apj, 838, 151

\bibitem[{{Turner} {et~al.}(2014){Turner}, {Fromang}, {Gammie}, {Klahr},
  {Lesur}, {Wardle}, \& {Bai}}]{2014prpl.conf..411T}
{Turner}, N.~J., {Fromang}, S., {Gammie}, C., {et~al.} 2014, Protostars and
  Planets VI, 411

\bibitem[{{Ulrich}(1976)}]{1976ApJ...210..377U}
{Ulrich}, R.~K. 1976, \apj, 210, 377

\bibitem[{{Wada} {et~al.}(2009){Wada}, {Tanaka}, {Suyama}, {Kimura}, \&
  {Yamamoto}}]{2009ApJ...702.1490W}
{Wada}, K., {Tanaka}, H., {Suyama}, T., {Kimura}, H., \& {Yamamoto}, T. 2009,
  \apj, 702, 1490

\bibitem[{{Wada} {et~al.}(2011){Wada}, {Tanaka}, {Suyama}, {Kimura}, \&
  {Yamamoto}}]{2011ApJ...737...36W}
{Wada}, K., {Tanaka}, H., {Suyama}, T., {Kimura}, H., \& {Yamamoto}, T. 2011,
  \apj, 737, 36

\bibitem[{{Weidenschilling}(1977)}]{1977Ap&SS..51..153W}
{Weidenschilling}, S.~J. 1977, \apss, 51, 153

\bibitem[{{Weidenschilling}(1980)}]{1980Icar...44..172W}
{Weidenschilling}, S.~J. 1980, \icarus, 44, 172

\end{thebibliography}

\end{document}